\documentclass[journal]{IEEEtran}
\usepackage{amsmath,amsfonts,amssymb}
\usepackage{algorithmic}
\usepackage{array}
\usepackage[caption=false,font=normalsize,labelfont=sf,textfont=sf]{subfig}
\usepackage{textcomp}
\usepackage{stfloats}
\usepackage{url}
\usepackage{verbatim}
\usepackage{graphicx}
\usepackage{cite}
\usepackage{tikz}
\usepackage{orcidlink}
\usetikzlibrary{patterns}
\newcommand{\vect}[1]{\boldsymbol{#1}}
\begin{document}

\title{Damping Oscillations in a Spherical Pendulum Inclinometer\\ Using Vector-Based Control}

\author{Fernando Capes \orcidlink{0009-0007-7078-6773}, Mikael Andreas Bianchi \orcidlink{0000-0001-6510-9275}, Roberto Gardenghi \orcidlink{0000-0002-2921-1210}, Manuel Alò \orcidlink{0009-0006-5744-929X} 

\thanks{F. Capes and M. Alò contributed substantially to this work, with F. Capes leading the modeling and experimental activities, and M. Alò contributing to system design and manuscript preparation.}
\thanks{Fernando Capes was a Master’s student at SUPSI ISEA, Lugano, TI, Switzerland}
\thanks{Mikael Andreas Bianchi, head of Mechatronic systems laboratory at SUPSI ISEA, Lugano, TI, Switzerland {\small\url{mikael.bianchi@supsi.ch}}}
\thanks{Roberto Gardenghi, Adjunct Professor and head of Applied Photonics and Optoelectronics Area at SUPSI ISEA, Lugano, TI, Switzerland {\small\url{roberto.gardenghi@supsi.ch}}}
\thanks{Manuel Alò, researcher at SUPSI ISEA, Lugano, TI, Switzerland {\small\url{manuel.alo@supsi.ch}}}
\thanks{This work has been submitted to the IEEE for possible publication. Copyright may be transferred without notice, after which this version may no longer be accessible.}
}

\maketitle

\begin{abstract}
This paper addresses disturbance-induced oscillations in high-precision pendulum-based inclinometers, which reduce measurement availability despite the long-term stability of gravity-referenced sensing. To actively suppress these oscillations, a contactless six-coil electromagnetic actuation system is developed, together with a control framework that generates, in real time, the planar damping force required by the controller despite the nonlinear, unilateral, and bounded nature of magnetic actuation. The proposed approach combines a control-oriented nonlinear model of the spherical pendulum, observer-based feedback control, and a constrained force-allocation scheme that maps the desired vector force into feasible coil currents. Experimental validation on a prototype demonstrates attenuation of the dominant oscillatory mode up to 31.11 dB and a marked reduction in transient duration, with the 50\% decay time decreasing from 11.10 s without control to 1.01 s at the most favorable operating point. The results also highlight a trade-off between transient speed and residual oscillation, demonstrating both the effectiveness of the proposed damping strategy and its value as a practical design framework for high-precision inclinometer systems.
\end{abstract}


\section{Introduction}
    \IEEEPARstart{D}{}igital inclinometers based on MEMS accelerometers are widely adopted for their compactness, low cost, and ease of integration. However, their long-term accuracy and stability are fundamentally limited by sensor-level effects such as temperature-dependent bias, mechanical stress, and aging-induced drift \cite{9955550,9335933,9171352}. Although compensation strategies have been proposed, residual drift remains a critical issue in applications requiring sustained high precision \cite{s150511222, PRIKHODKO2013517, jsan7030030}. For applications demanding microradian-level resolution, such as structural health monitoring of dams, pendulum-based inclinometers offer a compelling alternative. By relying on gravity as a stable physical reference, these systems inherently reduce drift and enable material and structural optimization to mitigate temperature-induced effects \cite{Yang2025HighPrecision, Taylor2025UltraSensitive, Cui2018HighResolution}. Despite their high precision, pendulum-based sensors suffer from low intrinsic damping because dissipative effects are limited at small amplitudes. Residual damping is mainly associated with internal friction in the suspension, leading to persistent oscillations under external disturbances and increased settling time \cite{Dolfo2016PendulumDamping}. Passive damping strategies can increase dissipation, but viscous approaches may introduce mechanical coupling, thermal noise, or equilibrium perturbations, whereas eddy-current damping is contactless but provides limited authority at low oscillation velocities \cite{Plissi2004EddyCurrentDamping, Chen2013ECD}. Furthermore, the literature on high-precision inclinometers has largely focused on static accuracy, with comparatively limited attention to dynamic behavior under disturbance-induced oscillations \cite{LaserInclinometer}. Active damping therefore represents a promising alternative to improve transient response, at the cost of increased control and actuation complexity \cite{LSTM_RNN_Vibration, ShakingTableControl}. Electromagnetic actuation is particularly suited for high-precision systems, as it enables contactless force generation and avoids mechanical nonlinearities such as friction and backlash \cite{MagLevReview, MagLevPositioning}. Its effectiveness in increasing dynamic stiffness and suppressing vibrations has been widely demonstrated in magnetic bearings and positioning platforms \cite{Knospe2007AMBMachining, Trumper1997LinearizingMagneticSuspension}. However, the practical use of electromagnetic actuation requires explicit treatment of force generation, as magnetic forces are inherently nonlinear, operating-point dependent, and strongly influenced by geometry and field distribution \cite{Abbott2007SoftMagneticBodies, Yu2025HarmonicMB}. Recent multi-coil systems further show that distributed force generation can be exploited for complex motion tasks, albeit with configuration-dependent behavior \cite{Lee2025MEMA}. In pendulum-based inclinometers, these considerations make the joint design of actuator configuration, force modeling, and control strategy particularly critical, since oscillations must be damped without disturbing equilibrium or degrading measurement accuracy. From a control viewpoint, pendulum dynamics have been extensively studied, from planar to spherical formulations, in which additional degrees of freedom introduce strong nonlinear coupling and complex stability behavior \cite{Shen2004_3DPendulum, Miles1962SphericalPendulum}. However, much of the existing literature relies on idealized models, whereas practical implementations must operate under actuator saturation, nonlinear force characteristics, and limited bandwidth \cite{ShakingTableControl, MSMAActuator}. Against this background, this work investigates active damping under realistic actuation constraints using a vectorial electromagnetic system specifically designed for this application. The proposed approach combines a physics-based modeling framework with an optimization-based force allocation strategy, enabling effective oscillation suppression while operating within the actuator's physical limits and preserving measurement integrity.

\section{System description and problem formulation}\label{sec:system-description-problem-formulation}
    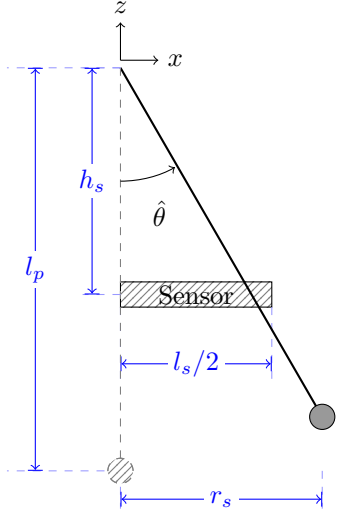
\begin{figure}[!t]
    \centering
%
\newcommand{\nvar}[2]{%
    \newlength{#1}
    \setlength{#1}{#2}
}

\newcommand{\lineann}[4][0.5]{%
    \begin{scope}[rotate=#2, blue,inner sep=2pt]
        \draw[dashed, blue!40] (0,0) -- +(0,#1)
            node [coordinate, near end] (a) {};
        \draw[dashed, blue!40] (#3,0) -- +(0,#1)
            node [coordinate, near end] (b) {};
        \draw[|<->|] (a) -- node[fill=white] {#4} (b);
    \end{scope}
}

\newcommand{\angann}[3]{%
\nvar{\ddxOne}{1.5cm}
    \begin{scope}[#3]
    \draw [->, shorten >=1pt] (\ddxOne,0pt) arc (0:#1:\ddxOne);
    \node at (#1/2:\ddxOne+14pt) {#2};
    \end{scope}
}

 \begin{tikzpicture}
    \pgfmathsetmacro{\Gvec}{1.5}
    \pgfmathsetmacro{\Pscale}{5/150}
    \pgfmathsetmacro{\Plen}{\Pscale*160}
    \pgfmathsetmacro{\PmassRadius}{\Pscale*5}
    \pgfmathsetmacro{\Psensoroffset}{\Pscale*100}
    \pgfmathsetmacro{\PsensorLen}{\Pscale*60}
    \pgfmathsetmacro{\PsensorWidth}{\Pscale*10}    
    \pgfmathsetmacro{\myAngle}{30}
    \pgfmathsetmacro{\Psin}{\Plen*sin(\myAngle)}

    \coordinate (centro) at (0,0);
    \begin{scope}[shift=(-90:\Psensoroffset-\PsensorWidth/2), rotate=0]
        \filldraw [pattern=north east lines,pattern color=gray] (0,0) rectangle++ (\PsensorLen,\PsensorWidth) node[pos=.5] {Sensor};
        \lineann[-1.0]{0}{\PsensorLen}{$l_s/2$}

    \end{scope}
    
    \begin{scope}[shift=(0:0), rotate=-90]
        \draw[thick] (centro) -- ++(\myAngle:\Plen) coordinate (mass);
        \draw[dashed,gray,-] (centro) -- ++(-0:\Plen) coordinate (zeroPos);
        \angann{\myAngle}{$\hat{\theta}$}{black} 
        \lineann[-1.5]{0}{\Plen}{$l_p$}
        \lineann[-0.5]{0}{\Psensoroffset-\PsensorWidth}{$h_s$}
    \end{scope}

    \filldraw [fill=black!40,draw=black] (mass) circle[radius=\PmassRadius];
    \begin{scope}[shift=(-90:\Plen), rotate=0]
        \lineann[-0.5]{0}{\Psin}{$r_s$}
    \end{scope}    
    
    \filldraw [fill=white,draw=gray,dashed] (zeroPos) circle[radius=\PmassRadius];
    \filldraw [pattern=north east lines,pattern color=gray,draw=gray,dashed] (zeroPos) circle[radius=\PmassRadius];
    \draw[->] (0,0.1)--(0.5,0.1) node[right]{$x$};
    \draw[->] (0,0.1)--(0,0.6) node[above]{$z$};

\end{tikzpicture}
    \caption{Principle scheme of a Pendulum-Based Inclinometer where: $l_p$ is the length of the pendulum and $\theta$ is the angular position}\label{fig:pendulum-scheme}
    \end{figure}
    This section introduces the inclinometer architecture and the simplified dynamic model adopted for problem formulation and preliminary actuator sizing. The mechanical system consists of a metallic capsule with radius 20 mm and height 180 mm, housing the pendulum and the measurement electronics. The sensor operates by pendulum-cable triangulation. Two LEDs, sequentially activated at a fixed horizontal distance, project two shadows onto a pixel array used to infer the pendulum position. Fig.~\ref{fig:pendulum-scheme} shows a schematic of the pendulum and the sensor half-section in the observation plane. Given the pendulum geometry and the sensor field of view, the in-plane motion is assumed to remain within a circular workspace of radius 15 mm, corresponding to a worst-case angular excursion of approximately $\theta_{\max}=\pi/30$ rad. After an external disturbance, the response is expected to be dominated by oscillation near the natural frequency. For control-oriented problem formulation and preliminary actuator sizing, the in-plane motion is therefore approximated by the following one-dimensional model:
    \begin{equation}
        \ddot{\theta} = -\frac{g}{l_p}\sin(\theta)-\frac{c_p}{m_p\,l_p^2}\,\dot{\theta}+\frac{1}{m_p\,l_p^2}\,\tau_{p}
        \label{eq:simple-pendulum-equation}
    \end{equation}
    where \(g\) is the gravitational acceleration \([\mathrm{m/s^2}]\), \(l_p\) is the pendulum length \([\mathrm{m}]\), \(m_p\) is the pendulum mass \([\mathrm{kg}]\), \(c_p\) is the equivalent viscous torsional damping coefficient \([\mathrm{N\,m\,s/rad}]\), \(\theta\) is the angular displacement \([\mathrm{rad}]\), \(\dot{\theta}\) is the angular velocity \([\mathrm{rad/s}]\), \(\ddot{\theta}\) is the angular acceleration \([\mathrm{rad/s^2}]\), and \(\tau_p\) is the external torque \([\mathrm{N\,m}]\). Under the small-angle assumption, the corresponding natural frequency is $f_n = \frac{1}{2\,\pi}\sqrt{\frac{g}{l_p}} \approx 1.17\,\text{Hz}$. The design objective is to size the actuation system so that, under the worst-case condition, it achieves at least 20 dB attenuation of the dominant oscillatory component at $f_n$.

\section{Electromagnetic design} \label{Sec Electromagnetic design}
    This section presents the design and optimization of the electromagnetic actuator used for active damping of the spherical pendulum. The actuator is sized to provide sufficient damping authority while remaining sufficiently predictable for accurate force modeling, without perturbing the pendulum's equilibrium or measurement stability.
    \subsection{Actuator architecture and geometric constraints}
        \begin{figure}[!t]
        \centering
        \includegraphics[width=.8\columnwidth]{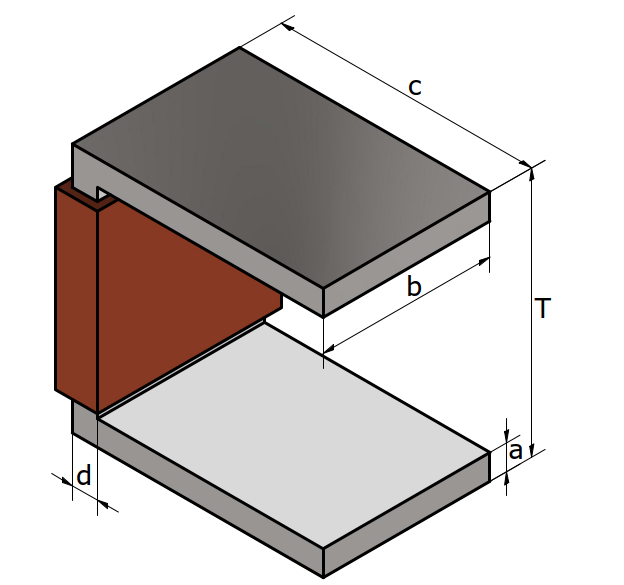}
        \caption{Geometry of the proposed C-shaped electromagnetic actuator, highlighting the main core dimensions used for the design and optimization process.}\label{fig:coil-dimensions}
        \end{figure}
        The actuator architecture was designed to balance force-generation capability with equilibrium preservation and mechanical constraints. To prevent residual static forces and parasitic stiffness that could bias the measurement, permanent magnets and hard magnetic materials were excluded from the pendulum mass. Consequently, the system operates exclusively via attractive forces. Finite Element Method (FEM) simulations identified a C-shaped core (Fig.~\ref{fig:coil-dimensions}) as the optimal compromise between flux concentration and usable force within the air gap. This geometry enhances flux guidance toward the pendulum workspace, ensuring a force distribution suitable for robust control design. To finalize the design within the mechanical envelope, the core dimensions were optimized in ANSYS Maxwell using a Sequential Mixed-Integer Nonlinear Programming (SMINLP) algorithm. The objective was to maximize the magnetic force at a 15~mm displacement, assuming a reference current density of 1~A/mm$^2$ to ensure a fair comparison among candidate geometries. The resulting optimal parameters are $T = 34$~mm, $a = 5$~mm, $b = 20$~mm, $c = 30$~mm, and $d = 5$~mm.

    \subsection{Energy-based actuator sizing and layout selection}\label{Sec Energy balance analysis}
        \begin{figure}[!t]
        \centering
        \includegraphics[width=\columnwidth]{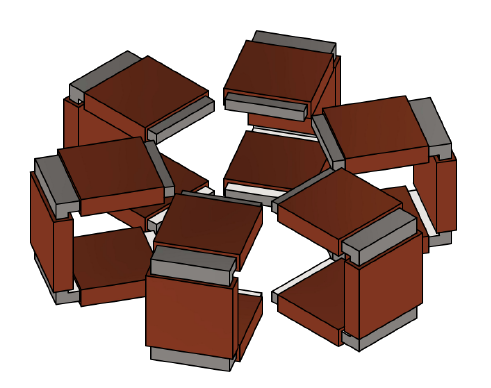}
        \caption{Evaluated actuator layouts and final six-coil configuration.}
        \label{fig:actuator-configurations}
        \end{figure}
        To complement the geometric optimization, a preliminary energy-based analysis was conducted to verify the damping authority of the actuator over the expected operating range. This assessment uses a simplified one-dimensional pendulum model as a conservative feasibility criterion for sizing, rather than for closed-loop performance prediction. The actuation system is designed to meet the 20~dB attenuation requirement for the dominant oscillatory component at $f_n$. Considering the worst-case initial amplitude $\theta_{\max}$ defined previously, the target residual amplitude is set to $\theta_{\mathrm{tar}} = \theta_{\max}/10$. Based on the mechanical energy $E$ of the pendulum:
        \begin{equation}
            E = \frac{1}{2} m_p l_p^2 \dot{\theta}^2 + m_p g l_p (1-\cos\theta).
            \end{equation}
        the energy $\Delta E_{\mathrm{req}}$ that must be dissipated within a single oscillation cycle to reach $\theta_{\mathrm{tar}}$ is:
        \begin{equation}
            \Delta E_{\mathrm{req}} = m_p g l_p \left(\cos\theta_{\mathrm{tar}} - \cos\theta_{\max}\right).
        \end{equation}
        The work $W_{\mathrm{ext}}$ extracted by the actuator must satisfy the feasibility condition $W_{\mathrm{ext}} \geq \Delta E_{\mathrm{req}}$. In a planar approximation, $W_{\mathrm{ext}}$ is defined as the line integral of the magnetic force $f(\theta,I)$ along the pendulum trajectory:
        \begin{equation}
            W_{\mathrm{ext}} = -\oint f(\theta,I)\,l_p\cos\theta\,d\theta.
        \end{equation}
        The extracted-work formulation was used as a preliminary actuator-sizing criterion by assuming, for each swing direction, that the pendulum motion occurred between the two adjacent coils able to provide a dissipative action. Since each electromagnet is purely attractive, the work over one cycle was obtained by summing the two half-cycle contributions, selecting in each half-cycle the coil whose horizontal force opposed the pendulum velocity. This criterion was then evaluated numerically over the full pendulum workspace for different actuator layouts. Using ANSYS Maxwell, the magnetic force generated by the candidate layouts was simulated for different pendulum positions and azimuthal swing orientations. For each configuration, the force component contributing to dissipation was interpolated over the workspace and numerically integrated over one oscillation cycle. Three layouts were considered, composed of 3, 4, and 6 coils equally spaced by $120^\circ$, $90^\circ$, and $60^\circ$, respectively. To compare the candidate layouts, the ratio:
        \begin{equation}
            \eta = \frac{\Delta E_{\mathrm{req}}}{W_{\mathrm{ext}}}
        \end{equation}
        was used as a feasibility index, with $\eta \leq 0.91$ considered acceptable to account for numerical approximation and modeling uncertainty. In this way, each layout was assessed over the admissible set of initial oscillation directions and amplitudes, rather than along a single favorable trajectory. Only the six-coil configuration satisfied the energy-based damping requirement over the full pendulum workspace, leading to the final design with six evenly spaced C-shaped electromagnets on a circumference of radius 50 mm, as shown in Fig.~\ref{fig:actuator-configurations}.

    \subsection{Control-oriented force model of the actuator}\label{Sec Control-oriented force model of the actuator}

        \subsubsection{Force--position relation}\label{sec Force-Position relation}

        \begin{figure}[!t]
        \centering
        \includegraphics[width=\columnwidth]{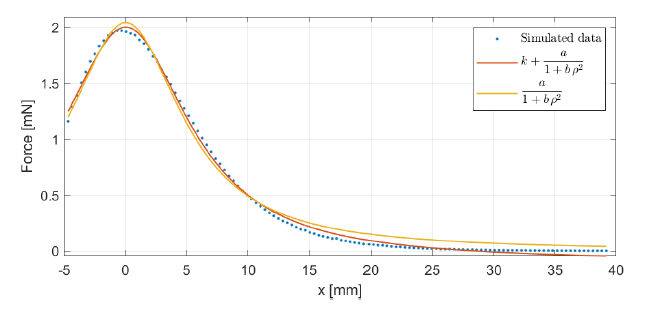}
        \caption{Fitted force function.}
        \label{fig:force-position-fit}
        \end{figure}

        For control design, a compact parametric model of the actuator force is required. Although analytical models for magnetic force estimation are available in the literature \cite{Abbott2007SoftMagneticBodies}, they are often cumbersome to use in practical design workflows. Here, a control-oriented model is identified from numerical simulations of a single active actuator operated at a reference excitation level. In this subsection, \(f\) denotes the magnitude of the attractive force projected onto the actuator plane. A widely used approximation for electromagnetic force modeling expresses the force magnitude as a second-order rational function of position \cite{Trumper1997LinearizingMagneticSuspension,9782635}. In Cartesian coordinates, this model can be written as
        \begin{equation}
            f(x,y)=\frac{a}{1+bx^2+cy^2},
        \end{equation}
        where \(a\) is the peak force amplitude, while \(b\) and \(c\) determine the force decay along the \(x\)- and \(y\)-directions, respectively. By introducing the polar transformation \(x=\rho\cos\varphi\), \(y=\rho\sin\varphi\), the model becomes
        \begin{equation}
            f(\rho,\varphi)=\frac{a}{1+b\rho^2\cos^2\varphi+c\rho^2\sin^2\varphi}.
        \end{equation}
        Since the simulated force surface exhibited an approximately isotropic decay in the considered operating region, the simplification \(b=c\) was adopted, yielding
        \begin{equation}
            f(\rho)=\frac{a}{1+b\rho^2}.
            \label{eq:force-position}
        \end{equation}
        Fig.~\ref{fig:force-position-fit} compares the simulated force data with the fitting result obtained from \eqref{eq:force-position}.

        \subsubsection{Force--current relation}\label{sec Force-Current relation}
        \begin{figure}[!t]
            \centering
            \includegraphics[width=\columnwidth]{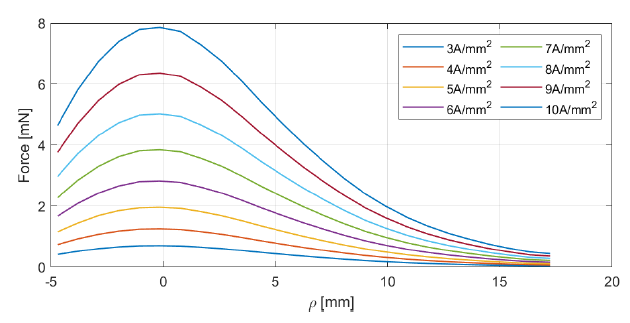}
            \caption{Force variation with current as parameter.}
            \label{fig:force-current-variation}
        \end{figure}
        After identifying the force--position relationship at a reference excitation level, the dependence on current was analyzed through additional simulations. As shown in Fig.~\ref{fig:force-current-variation}, the peak force varies quadratically with the excitation current. Accordingly, the peak amplitude can be written as \(a(I)=\alpha I^2\), and \eqref{eq:force-position} becomes
        \begin{equation}
            f(\rho,I)=\frac{\alpha I^2}{1+b\rho^2},
            \label{eq:force-current}
        \end{equation}
        where \(\alpha\) is the proportionality coefficient relating the peak force amplitude to the square of the coil current. Equation~\eqref{eq:force-current} provides the control-oriented scalar model used in the following sections for actuator characterization and real-time force allocation.

\section{Spherical pendulum modeling}\label{sec:spherical-pendulum-modeling}
    \begin{figure}[!t]
    \centering 
    \includegraphics[width=\columnwidth]{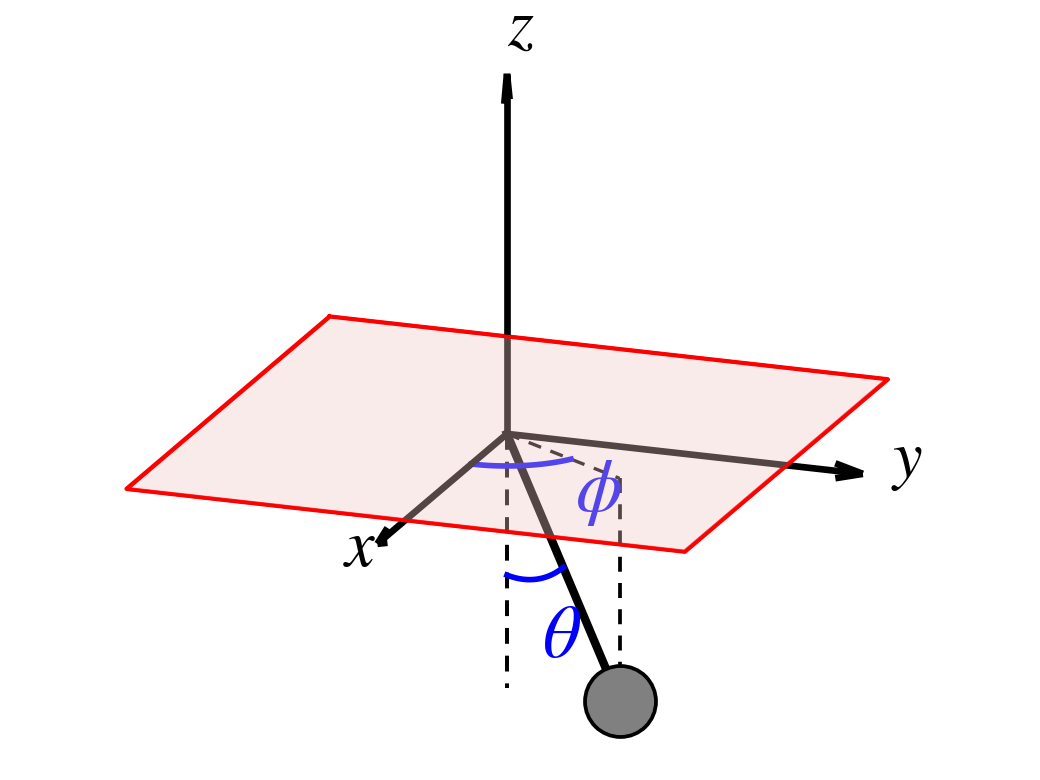}
    \caption{Spherical pendulum parameterization in spherical coordinates.}
    \label{fig:spherical-pendulum-spherical-coordinates}
    \end{figure}
    The inclinometer is modeled as a spherical pendulum, i.e., a point mass \(m_p\) constrained to move on a sphere of fixed radius \(l_p\). Let \(\mathbf{q} := (\theta\; \phi)^T\) denote the generalized coordinates, where \(\theta\) is the polar angle and \(\phi\) is the azimuth angle, as shown in Fig.~\ref{fig:spherical-pendulum-spherical-coordinates}. The Cartesian position of the pendulum mass is
    \begin{equation}
        \mathbf{x} :=
        \begin{pmatrix}
        x\\
        y\\
        z
        \end{pmatrix}
        =
        \begin{pmatrix}
        l_p \sin\theta \cos\phi\\
        l_p \sin\theta \sin\phi\\
        -l_p \cos\theta
        \end{pmatrix}.
        \label{eq:cartesian-to-spherical}
    \end{equation}
    Differentiating \eqref{eq:cartesian-to-spherical} with respect to time yields
    \begin{equation}
        \dot{\mathbf{x}} =
        \mathbf{J}(\mathbf{q})\,\dot{\mathbf{q}},
        \label{eq:velocity-jacobian}
    \end{equation}
    where
    \begin{equation}
        \mathbf{J}(\mathbf{q}) =
        \begin{pmatrix}
        l_p \cos\phi \cos\theta & -l_p \sin\phi \sin\theta\\
        l_p \sin\phi \cos\theta & \phantom{-}l_p \cos\phi \sin\theta\\
        l_p \sin\theta & 0
        \end{pmatrix}
        \label{eq:jacobian}
    \end{equation}
    is the Jacobian matrix of the spherical parameterization. From \eqref{eq:velocity-jacobian}--\eqref{eq:jacobian}, the squared speed is
    \begin{equation}
        \dot{\mathbf{x}}^\top \dot{\mathbf{x}}
        =
        l_p^2\left(\dot{\theta}^2 + \sin^2\theta\,\dot{\phi}^2\right).
        \label{eq:speed-squared}
    \end{equation}
    Accordingly, the kinetic and potential energies are
    \begin{equation}
        \begin{aligned}
            T &= \frac{1}{2} m_p l_p^2 \left(\dot{\theta}^2 + \sin^2\theta\,\dot{\phi}^2\right),
            &\quad
            U &= m_p g l_p (1-\cos\theta),
        \end{aligned}
        \label{eq:mechanical-energies}
    \end{equation}
    where the zero-potential reference is chosen at the lower equilibrium position. Dissipation is modeled through the Rayleigh function
    \begin{equation}
        \mathcal{R}
        =
        \frac{1}{2} c_p l_p^2
        \left(
        \dot{\theta}^2 + \sin^2\theta\,\dot{\phi}^2
        \right),
        \label{eq:rayleigh}
    \end{equation}
    where \(c_p\) is an equivalent viscous damping coefficient. Let \(\mathbf{f} := \begin{pmatrix} f_x & f_y & f_z \end{pmatrix}^{\top}\) be an external Cartesian force applied to the pendulum mass. The corresponding generalized forces are obtained from the principle of virtual work as
    \begin{equation}
    \begin{pmatrix}
    Q_\theta\\
    Q_\phi
    \end{pmatrix}
    =
    \mathbf{J}^{\top}(\mathbf{q})\,\mathbf{f}.
    \label{eq:external-force-relation}
    \end{equation}
    Using the Lagrangian \(\mathcal{L}=T-U\), the equations of motion follow from
    \begin{equation}
    \frac{d}{dt}\!\left(\frac{\partial \mathcal{L}}{\partial \dot{q}_i}\right)
    -
    \frac{\partial \mathcal{L}}{\partial q_i}
    +
    \frac{\partial \mathcal{R}}{\partial \dot{q}_i}
    =
    Q_i,
    \qquad i\in\{\theta,\phi\}.
    \label{eq:euler-lagrange}
    \end{equation}
    After straightforward algebra, the nonlinear dynamics of the spherical pendulum can be written as
    \begin{equation}
    \left\{
    \begin{aligned}
    \ddot{\theta}
    &=
    -\dfrac{g}{l_p}\sin\theta
    +\dot{\phi}^2\sin\theta\cos\theta
    -\dfrac{c_p}{m_p}\dot{\theta}
    \\
    &\quad
    +\dfrac{f_z\sin\theta + f_x\cos\theta\cos\phi + f_y\cos\theta\sin\phi}{l_p m_p}
    \\[8pt]
    \ddot{\phi}
    &=
    -2\dot{\theta}\dot{\phi}\cot\theta
    -\dfrac{c_p}{m_p}\dot{\phi}
    +\dfrac{-f_x\sin\phi + f_y\cos\phi}{l_p m_p \sin\theta}.
    \end{aligned}
    \right.
    \label{eq:spherical-pendulum-model}
    \end{equation}
    For the intended application, it is convenient to express the dynamics in a force-driven form. The total Cartesian force acting on the pendulum is decomposed as
    \begin{equation}
    \mathbf{f} = \mathbf{f}_g + \mathbf{f}_a,
    \label{eq:force-decomposition}
    \end{equation}
    where \(\mathbf{f}_g\) is the gravity vector expressed in the chosen reference frame, and \(\mathbf{f}_a\) is the actuator force. With this formulation, the effect of gravity is represented as an external vector input rather than being embedded exclusively in the potential term. This representation is particularly useful for inclinometer modeling, since it allows the apparent gravity direction to vary with the sensor orientation while keeping the equations in a compact vector-input form. Equivalently, by removing the explicit gravitational term from \eqref{eq:spherical-pendulum-model} and including gravity directly in \(\mathbf{f}\), the model becomes
    \begin{equation}
    \left\{
    \begin{aligned}
    \ddot{\theta}
    &=
    \dot{\phi}^2\sin\theta\cos\theta
    -\dfrac{c_p}{m_p}\dot{\theta}
    \\
    &\quad
    +\dfrac{f_z\sin\theta + f_x\cos\theta\cos\phi + f_y\cos\theta\sin\phi}{l_p m_p}
    \\[8pt]
    \ddot{\phi}
    &=
    -2\dot{\theta}\dot{\phi}\cot\theta
    -\dfrac{c_p}{m_p}\dot{\phi}
    +\dfrac{-f_x\sin\phi + f_y\cos\phi}{l_p m_p \sin\theta},
    \end{aligned}
    \right.
    \label{eq:non-rotated-system-model}
    \end{equation}
    where \(\mathbf{f}\) now denotes the total external force, including gravity.

    \subsection{Coordinate regularization through reference-frame rotation}\label{subsec:frame-rotation}
        \begin{figure}[!t]
        \centering
        \includegraphics[width=\columnwidth]{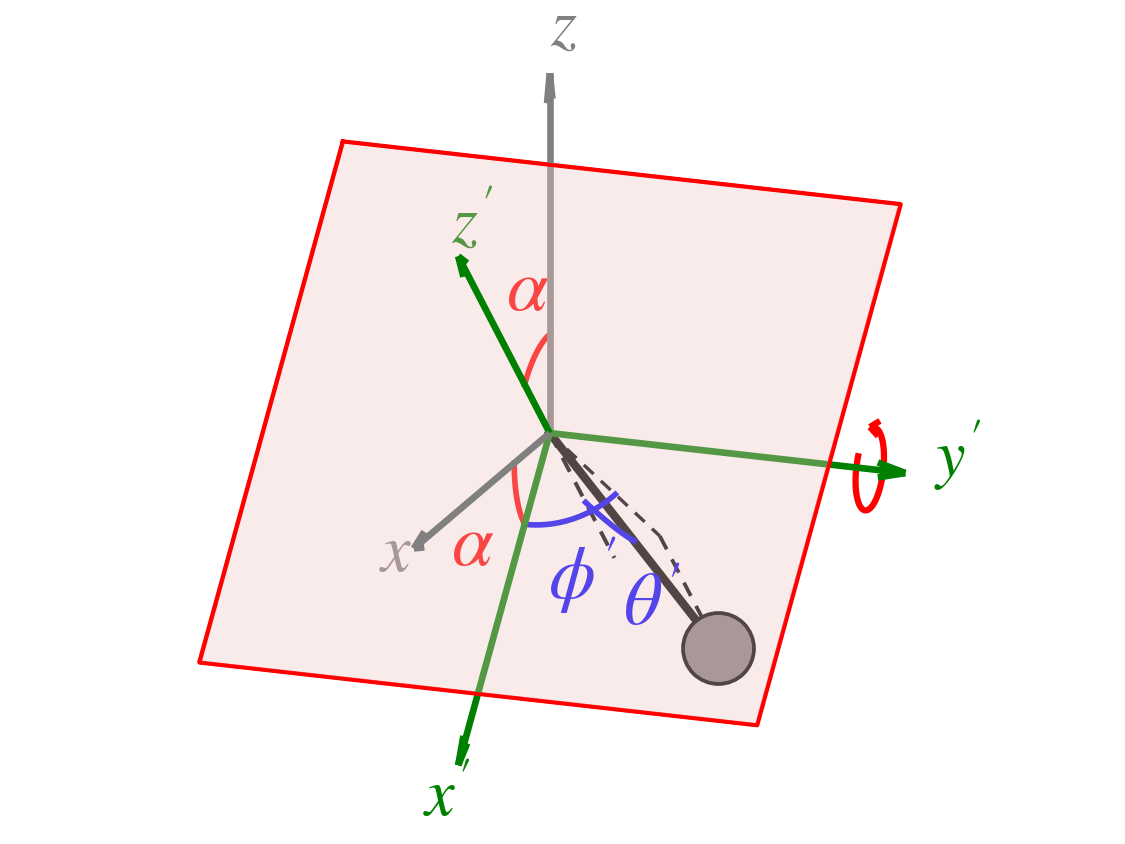}
        \caption{Rotated reference frame used to keep the operating region away from singular configurations of the spherical-coordinate model.}
        \label{fig:spherical-pendulum-rotated-references}
        \end{figure}
        A well-known limitation of spherical coordinates is the presence of singular configurations at \(\theta=0\) and \(\theta=\pi\), where the azimuth angle \(\phi\) becomes undefined and the dynamics in \eqref{eq:non-rotated-system-model} contain terms proportional to \(1/\sin\theta\). These singularities degrade numerical conditioning and complicate both linearization and controller design. Although global nonsingular representations such as quaternions may be adopted, they are unnecessary here because the pendulum operates in a narrow angular range. To regularize the model within the operating region, a fixed rotation of the reference frame is introduced. Let \(\mathcal{F}\) be the original inertial frame and \(\mathcal{F}'\) a frame obtained by rotating \(\mathcal{F}\) about the \(y\)-axis by an angle \(\alpha\). The Cartesian coordinates and force vector expressed in the rotated frame are
        \begin{equation}
        \mathbf{x}' = \mathbf{R}_y(\alpha)\,\mathbf{x},
        \qquad
        \mathbf{f}' = \mathbf{R}_y(\alpha)\,\mathbf{f},
        \label{eq:rotated-position-force}
        \end{equation}
        with
        \begin{equation}
        \mathbf{R}_y(\alpha) =
        \begin{pmatrix}
        \cos\alpha & 0 & \sin\alpha\\
        0 & 1 & 0\\
        -\sin\alpha & 0 & \cos\alpha
        \end{pmatrix}.
        \label{eq:rotation-matrix}
        \end{equation}
        The same spherical-pendulum model can then be written in the rotated frame by replacing \((\theta,\phi,\mathbf{f})\) with \((\theta',\phi',\mathbf{f}')\). In particular, the gravity vector in \(\mathcal{F}'\) becomes
        \begin{equation}
        \mathbf{f}'_g
        =
        \mathbf{R}_y(\alpha)
        \begin{pmatrix}
        0\\
        0\\
        -m_p g
        \end{pmatrix}
        =
        \begin{pmatrix}
        - m_p g \sin\alpha\\
        0\\
        - m_p g \cos\alpha
        \end{pmatrix}.
        \label{eq:rotated-gravity}
        \end{equation}
        For the choice \(\alpha=\pi/4\), the static equilibrium of the pendulum in the rotated frame is shifted away from the singular region and is given by
        \begin{equation}
        \theta_0' = \frac{\pi}{4},
        \qquad
        \phi_0' = \pi.
        \label{eq:rotated-equilibrium}
        \end{equation}
        This shift keeps the entire operating region sufficiently far from the poles of the spherical parameterization, thereby improving numerical robustness while preserving a compact two-angle representation. The rotated model is therefore adopted in the remainder of the paper for linearization, state estimation, and control design.

\section{Control design}
    This section presents the linearized state-space model and observer-based formulation used for active damping over the limited operating range.
    \subsection{Linearization}
        The model is linearized about the equilibrium point via a Taylor expansion. As established in Subsection~\ref{subsec:frame-rotation}, the equilibrium configuration is identified at $\pi/4$. By considering only the force acting on the mass as the gravitational force, the system can be expressed in its linearized form as:
        \begin{equation}
            \dot{\mathbf{x}} = \vect{A}\,\mathbf{x} + \vect{B}\,\mathbf{u}
            \label{eq:linearized-state-space-model}
        \end{equation}
        \begin{equation*}
            \mathbf{x} = \begin{pmatrix}
                \theta'\\ \dot{\theta}'\\ \phi'\\ \dot{\phi}'
            \end{pmatrix},
            \qquad
            \vect{A} =
            \begin{bmatrix}
                0 & 1 & 0 & 0\\
                -\dfrac{g}{l_p} & -\dfrac{c_p}{m_p} & 0 & 0\\
                0 & 0 & 0 & 1\\
                0 & 0 & -\dfrac{g}{l_p} & -\dfrac{c_p}{m_p}
            \end{bmatrix}
        \end{equation*}
        \begin{equation*}
            \mathbf{u} = \begin{pmatrix}
                f_x\\ f_y\\ f_z
            \end{pmatrix},
            \qquad
            \vect{B} =
            \begin{bmatrix}
                0 & 0 & 0\\
                -\dfrac{1}{l_p\,m_p} & 0 & 0\\
                0 & 0 & 0\\
                0 & -\dfrac{\sqrt{2}}{l_p\,m_p} & 0
            \end{bmatrix}
        \end{equation*}
        where the initial state vector $\vect{x_0} = \begin{pmatrix}\pi/4&0&\pi&0\end{pmatrix}^T$ and the initial input vector $\vect{u_0} = \begin{pmatrix}0&0&g\,m_p\end{pmatrix}^T$. The results indicate that for small oscillations, the spherical pendulum can be approximated as two independent mass-spring-damper systems, simplifying controller design. Moreover, near equilibrium, the vertical force has no effect on the system's acceleration, consistent with the assumption of a rigid, fixed-length pendulum rod.
        
    \subsection{Velocity estimation}
        Given the slow dynamics of the inclination measurement system, which would not allow for accurate velocity estimation, a state observer is introduced. A Luenberger observer was deemed unsuitable for this application, as any deviation of the gravitational direction (representing the system's inclination) from the perpendicular position to the plane would lead to incorrect velocity estimation. To address this issue, a Disturbance Observer (DOB) is proposed \cite{Observer}, an approach shown to improve control precision in systems subject to unknown force disturbances \cite{10636094}. This approach extends the classical observer by incorporating a disturbance feedforward control, enhancing the system's robustness and ensuring reliable velocity estimation under varying inclination conditions, as demonstrated by systems using DOBs for motion stabilization under uncertainty \cite{10549974}. By defining $d_\theta$ and $d_\phi$ as the disturbance in both polar and azimuth angles, the extended observer system is:
        \begin{equation}
            \begin{pmatrix}
                \dot{\mathbf{x}} \\ \dot{\mathbf{d}}
            \end{pmatrix} =
            \underbrace{
                \begin{bmatrix}
                    \mathbf{A} & \mathbf{A_{DOB}} \\
                    \mathbf{0} & \mathbf{0}
                \end{bmatrix}
            }_{\mathbf{A}_{aug}}
            \cdot
            \begin{pmatrix}
                \mathbf{x} \\ \mathbf{d}
            \end{pmatrix}
            +
            \underbrace{
                \begin{bmatrix}
                    \mathbf{B} \\
                    \mathbf{0}
                \end{bmatrix}
            }_{\mathbf{B}_{aug}}
            \cdot        
            \mathbf{f}
            \label{eq:observer-augmented-model}
        \end{equation}
        where
        \begin{equation}
            \begin{aligned}
                \mathbf{A}_{DOB} &=
                \begin{bmatrix}
                    0 & 0 \\
                    -\frac{1}{l_p\,m_p} & 0 \\
                    0 & 0 \\
                    0 & -\frac{\sqrt{2}}{l_p\,m_p}
                \end{bmatrix}, 
                \qquad
                \mathbf{d} &=
                \begin{pmatrix}
                    d_\theta \\
                    d_\phi
                \end{pmatrix}.
            \end{aligned}
            \label{eq:dob-definitions}
        \end{equation}
        The newly introduced term in the $\mathbf{B}_{aug}$ matrix corresponds to the physical magnitude of the gravitational force, thereby incorporating its estimated effect into the observer model.

    \subsection{Velocity-based damping controller}\label{sec:damping-force-generation}
        Using the observer estimates, damping is implemented through a velocity-based feedback law. Since the inclinometer is intended for quasi-static inclination measurement, no position feedback is introduced. This prevents the controller from modifying the static equilibrium position and reduces the risk of biasing the measured inclination. Let \(\hat{\mathbf{x}}\) denote the state estimated by the observer. The velocity components used for feedback are selected as
        \begin{equation}
        \hat{\mathbf{v}}
        =
        \mathbf{S}_v\hat{\mathbf{x}},
        \qquad
        \mathbf{S}_v =
        \begin{bmatrix}
        0 & 1 & 0 & 0\\
        0 & 0 & 0 & 1
        \end{bmatrix}.
        \label{eq:velocity-selection}
        \end{equation}
        A virtual damping acceleration is then imposed as
        \begin{equation}
        \mathbf{a}_d = -\mathbf{K}_v\hat{\mathbf{v}},
        \qquad
        \mathbf{K}_v =
        \begin{bmatrix}
        k_\theta & 0\\
        0 & k_\phi
        \end{bmatrix},
        \label{eq:virtual-damping-law}
        \end{equation}
        where \(k_\theta\) and \(k_\phi\) are positive damping gains. Using the planar input map extracted from the linearized model, the corresponding reference force is computed as
        \begin{equation}
        \mathbf{f}_{\mathrm{ref}}
        =
        \mathbf{B}_p^{-1}\mathbf{a}_d,
        \qquad
        \mathbf{B}_p =
        \begin{bmatrix}
        -\dfrac{1}{l_p m_p} & 0\\[4pt]
        0 & -\dfrac{\sqrt{2}}{l_p m_p}
        \end{bmatrix}.
        \label{eq:reference-force-controller}
        \end{equation}
        The resulting \(\mathbf{f}_{\mathrm{ref}}\) represents the planar force that would produce the desired damping action in the linearized model. This force is then passed to the allocation stage, which accounts for unilateral actuation, current saturation, and the position-dependent force capability of the six-coil actuator.

    \subsection{Constrained force allocation for the six-coil actuator}\label{sec:force-allocation}
        \begin{figure}[!t]
        \centering
        \includegraphics[width=\columnwidth]{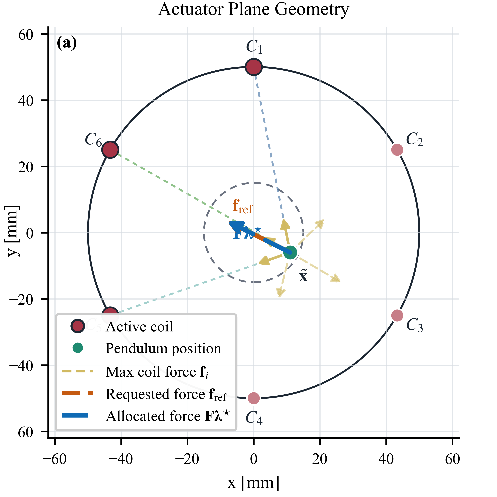}
        \caption{Geometric interpretation of the force-allocation problem. The green dot denotes the pendulum position projected onto the actuator plane, while the red dots indicate the coil centers. The yellow arrows represent the maximum force vectors generated by each coil at \(I_{\max}\), the red arrow is the reference force required by the controller, whereas the blue arrow shows the allocated force.}
        \label{fig:vector-forces-actuator-plane}
        \end{figure}
        The velocity-feedback controller provides a desired planar force \(\mathbf{f}_{\mathrm{ref}} \in \mathbb{R}^2\). However, the six-coil actuator cannot generate an arbitrary vector in the \(XY\) plane. Each electromagnet can only exert an attractive force toward its own center, and the force magnitude depends on both the applied current and the instantaneous pendulum position, as discussed in Section~\ref{Sec Control-oriented force model of the actuator}. Therefore, a force-allocation stage is required to map \(\mathbf{f}_{\mathrm{ref}}\) into six physically admissible coil currents. Let \(\tilde{\mathbf{x}} = (x\; y)^T\) be the projection of the pendulum position onto the actuator plane, and let \(\tilde{\mathbf{x}}_{c,i}\) denote the center of coil \(i\), with \(i=1,\dots,6\). For each coil, define the displacement vector \(\Delta \mathbf{x}_i = \tilde{\mathbf{x}}_{c,i} - \tilde{\mathbf{x}}\), its Euclidean norm \(\rho_i = \|\Delta \mathbf{x}_i\|_2\), and the corresponding unit direction \(\hat{\mathbf{u}}_i = \Delta \mathbf{x}_i / \rho_i\), so that \(\hat{\mathbf{u}}_i\) points from the pendulum toward coil \(i\). Using the force model in \eqref{eq:force-current}, the maximum force vector generated by coil \(i\) at the admissible current limit \(I_{\max}\) is
        \begin{equation}
        \bar{\mathbf{f}}_i(\tilde{\mathbf{x}})
        =
        f(\rho_i,I_{\max})\,\hat{\mathbf{u}}_i.
        \label{eq:max-coil-force-vector}
        \end{equation}
        Since the magnetic force scales with the square of the current, the normalized actuation variable is defined as
        \begin{equation}
        \lambda_i := \left(\frac{I_i}{I_{\max}}\right)^2,
        \qquad
        0 \le \lambda_i \le 1.
        \label{eq:lambda-definition}
        \end{equation}
        With this definition, \(\lambda_i=0\) corresponds to zero current, while \(\lambda_i=1\) corresponds to the maximum allowable current. Under the adopted control-oriented superposition assumption, the total force generated by the actuator is written as
        \begin{equation}
        \mathbf{f}_a(\tilde{\mathbf{x}},\boldsymbol{\lambda})
        =
        \sum_{i=1}^{6}\lambda_i\,\bar{\mathbf{f}}_i(\tilde{\mathbf{x}})
        =
        \mathbf{F}(\tilde{\mathbf{x}})\boldsymbol{\lambda},
        \label{eq:allocated-force}
        \end{equation}
        where \(\mathbf{F}(\tilde{\mathbf{x}}) := [\bar{\mathbf{f}}_1(\tilde{\mathbf{x}})\ \cdots\ \bar{\mathbf{f}}_6(\tilde{\mathbf{x}})] \in \mathbb{R}^{2\times 6}\) collects the six maximum coil-force vectors.
        For a fixed pendulum position, the set of achievable planar forces is therefore
        \begin{equation}
        \mathcal{P}(\tilde{\mathbf{x}})
        =
        \left\{
        \mathbf{F}(\tilde{\mathbf{x}})\boldsymbol{\lambda}
        \,:\,
        \boldsymbol{\lambda}\in[0,1]^6
        \right\}.
        \label{eq:feasible-force-set}
        \end{equation}
        If \(\mathbf{f}_{\mathrm{ref}} \in \mathcal{P}(\tilde{\mathbf{x}})\), the requested force can be reproduced exactly. In general, however, exact reproduction is not always possible because of unilateral actuation and current saturation. For this reason, the force-allocation problem is formulated as the following bounded least-squares optimization:
        \begin{equation}
        \begin{aligned}
        \min_{\boldsymbol{\lambda}} \quad &
        \frac{1}{2}
        \left\|
        \mathbf{F}(\tilde{\mathbf{x}})\boldsymbol{\lambda}
        -
        \mathbf{f}_{\mathrm{ref}}
        \right\|_2^2
        +
        \frac{\varepsilon}{2}\|\boldsymbol{\lambda}\|_2^2
        \\
        \text{s.t.}\quad &
        \mathbf{0}\leq\boldsymbol{\lambda}\leq\mathbf{1},
        \end{aligned}
        \label{eq:force-allocation-qp}
        \end{equation}
        where \(\varepsilon>0\) is a small regularization coefficient. The first term minimizes the force-tracking error, while the second penalizes actuation effort, improves numerical conditioning, and makes the problem strictly convex. The coefficient \(\varepsilon\) is chosen small so that the allocation remains dominated by force tracking. The Hessian matrix associated with \eqref{eq:force-allocation-qp} is
        \[
        \mathbf{H}(\tilde{\mathbf{x}})
        =
        \mathbf{F}^T(\tilde{\mathbf{x}})\mathbf{F}(\tilde{\mathbf{x}})
        +
        \varepsilon \mathbf{I}_6,
        \]
        which is symmetric positive definite for any \(\varepsilon>0\). Since the feasible set \([0,1]^6\) is convex and compact, problem \eqref{eq:force-allocation-qp} admits a unique solution \(\boldsymbol{\lambda}^\star\) for every pendulum position and every reference force. The corresponding coil currents are then recovered from \eqref{eq:lambda-definition} as
        \begin{equation}
        I_i = I_{\max}\sqrt{\lambda_i^\star},
        \qquad i=1,\dots,6.
        \label{eq:current-recovery}
        \end{equation}
        For real-time implementation, \eqref{eq:force-allocation-qp} can be solved online through a projected-gradient iteration:
        \begin{equation}
        \boldsymbol{\lambda}^{(k+1)}
        =
        \Pi_{[0,1]^6}
        \left[
        \boldsymbol{\lambda}^{(k)}
        -
        \alpha
        \left(
        \mathbf{H}\boldsymbol{\lambda}^{(k)}
        -
        \mathbf{F}^T\mathbf{f}_{\mathrm{ref}}
        \right)
        \right],
        \label{eq:projected-gradient-allocation}
        \end{equation}
        where \(\Pi_{[0,1]^6}(\cdot)\) denotes the Euclidean projection onto the hypercube \([0,1]^6\), i.e., component-wise saturation in the interval \([0,1]\). For any constant step size satisfying
        \[
        0 < \alpha < \frac{2}{\lambda_{\max}(\mathbf{H})},
        \]
        the iteration converges to the unique minimizer of \eqref{eq:force-allocation-qp}. This makes the proposed allocation strategy suitable for embedded implementation with limited computational resources. Fig.~\ref{fig:vector-forces-actuator-plane} provides a geometric interpretation of the allocation problem. For a given pendulum position, each coil generates a position-dependent force vector directed toward its center, and the allocator determines the bounded combination of these six vectors that best approximates the reference force required by the controller.

\section{Experimental validation}
\begin{figure}[!t]
        \centering
        \includegraphics[width=\columnwidth]{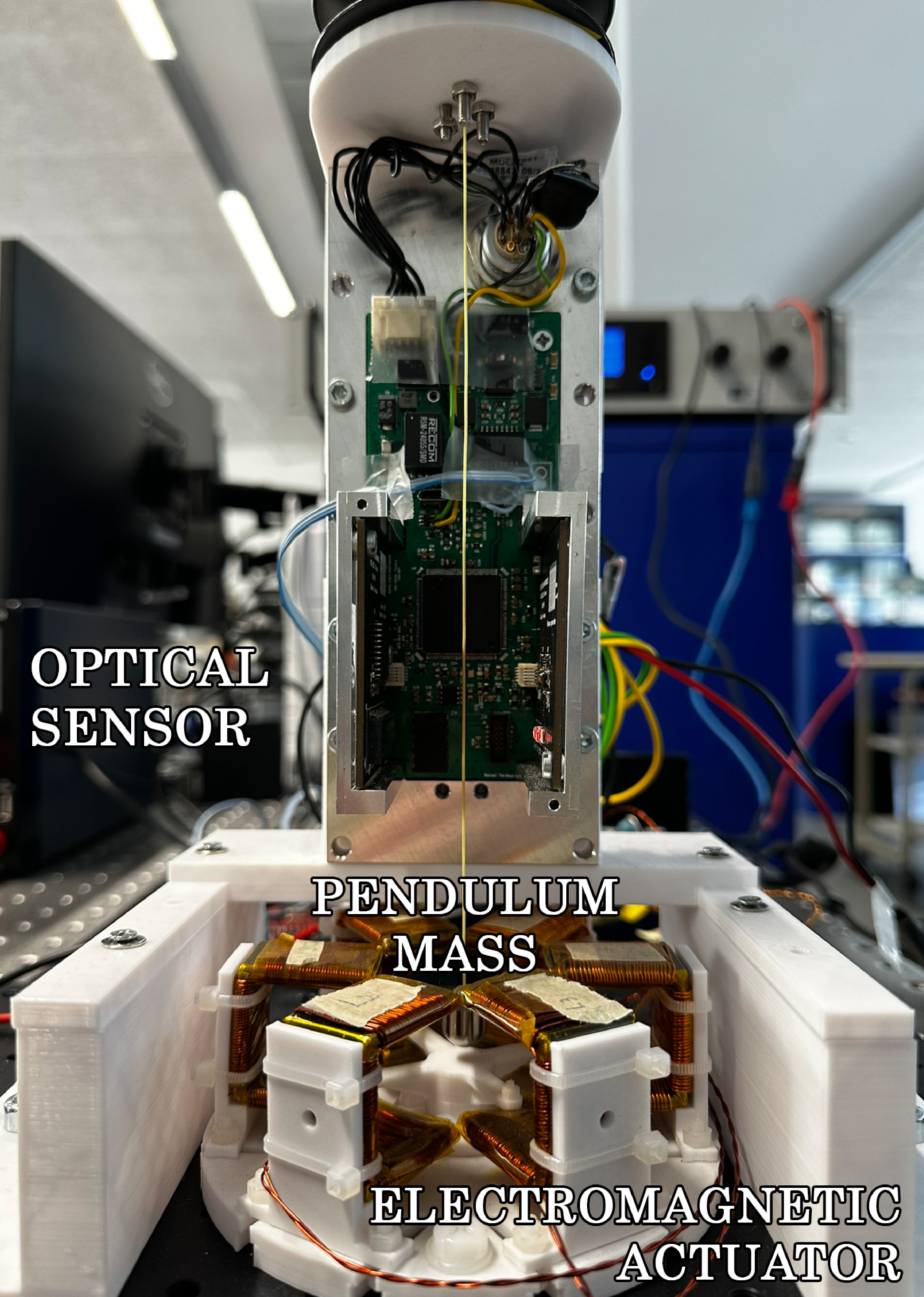}
        \caption{Experimental prototype used for validation. The setup integrates the optical subsystem for pendulum position measurement and the magnetic subsystem for active dynamic conditioning.}
        \label{fig:prototype-experimental-setup}
    \end{figure}

    The experimental campaign comprised preliminary parameter identification followed by closed-loop validation of the proposed damping strategy. The original optical readout was conceived for quasi-static operation; therefore, for the experimental campaign, the legacy filtering was bypassed, increasing the effective measurement update rate from about 77 ms to 7.7 ms and making real-time closed-loop validation feasible. Since only pendulum position was directly measured, observer-based feedback was used to reconstruct the velocity states required by the damping controller. The pendulum mass, length, and friction coefficient were identified from disturbance-induced transients through nonlinear grey-box estimation after offset correction and conversion to spherical coordinates, yielding a model fit error below 0.6\%. The actuator coefficients \(\alpha\) and \(b\) in \eqref{eq:force-current} were estimated from static equilibrium measurements under constant current through constrained optimization, with results consistent with magnetostatic simulations. The complete control architecture, including state estimation, feedback control, and force allocation, was implemented on a microcontroller. Fig.~\ref{fig:prototype-experimental-setup} shows the prototype used for experimental validation, highlighting both the optical subsystem for pendulum position measurement and the magnetic subsystem for active dynamic conditioning.

    \subsection{Closed-loop system validation}
        \begin{figure*}[!t]
            \centering
            \subfloat[]{\includegraphics[width=\columnwidth]{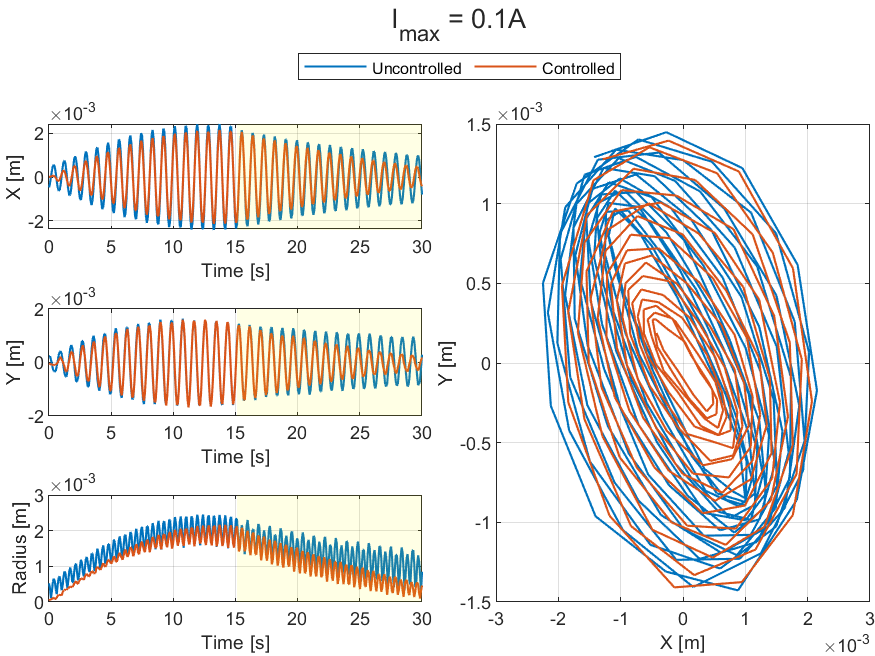}%
            \label{fig:imax_01}}
            \subfloat[]{\includegraphics[width=\columnwidth]{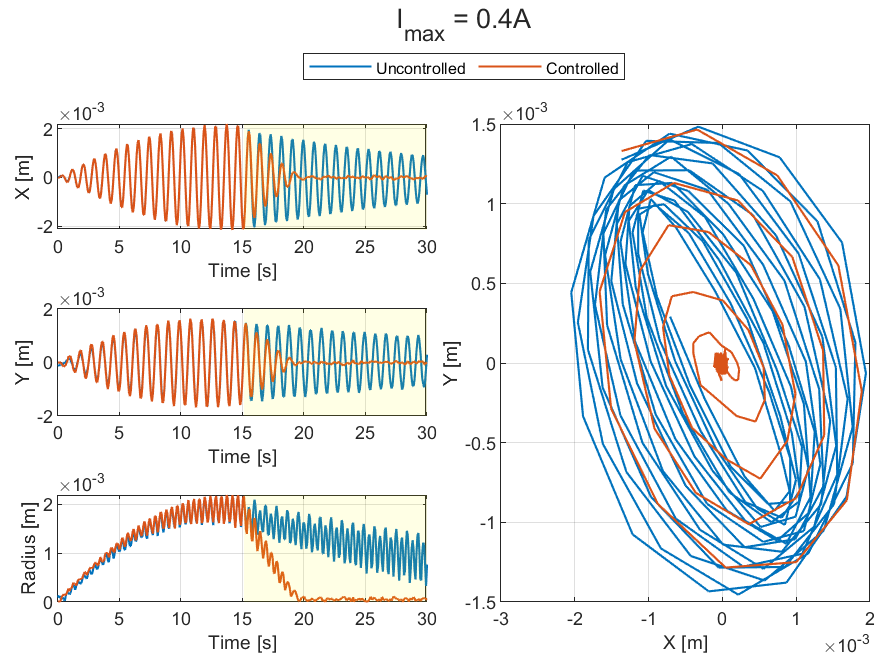}%
            \label{fig:imax_04}}

            \subfloat[]{\includegraphics[width=\columnwidth]{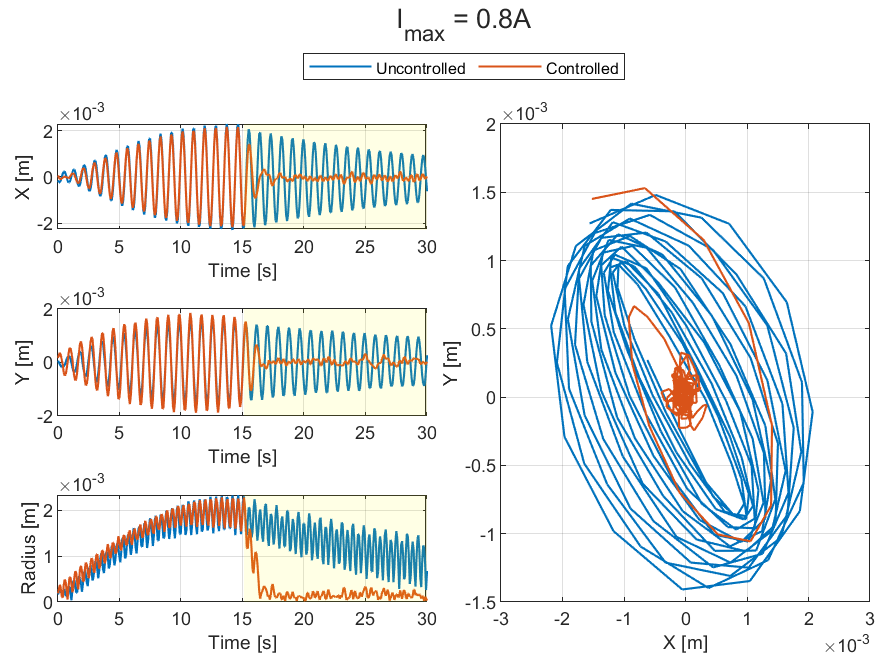}%
            \label{fig:imax_08}}
            \subfloat[]{\includegraphics[width=\columnwidth]{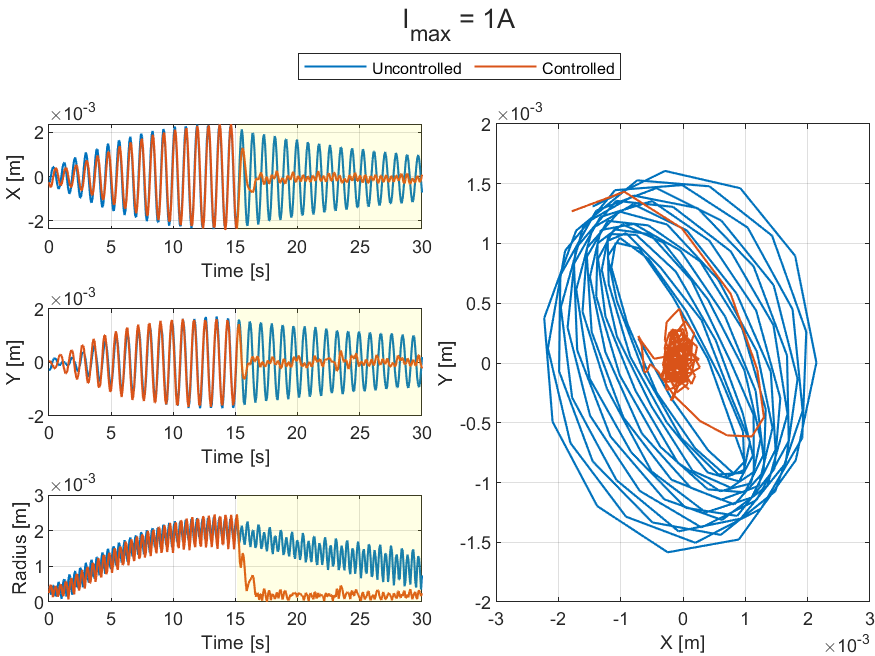}%
            \label{fig:imax_1}}
            \caption{Pendulum response under different maximum control currents. In each plot, the blue curve shows the response without control, while the red curve shows the response with control activated at half of the acquisition time (highlighted in yellow).}
            \label{fig:closed-loop-test-results}
        \end{figure*}
        Following the system identification phase, experimental tests were conducted to validate the closed-loop control performance. A repeatable disturbance was introduced by applying a controlled excitation to the coils, and the pendulum response was recorded with and without the control algorithm enabled. A video demonstrating the closed-loop damping performance is provided as supplementary material. Fig.~\ref{fig:closed-loop-test-results} provides a qualitative comparison of the system response for different values of the maximum allowable current. The plots show that the proposed controller is able to significantly reduce the oscillation amplitude even when the available actuation authority is strongly limited, i.e., when only a small electromagnetic force can be generated. Increasing the current limit leads to a faster suppression of the transient, at the expense of a higher residual oscillation level in the final part of the response. To quantitatively assess the effect of the maximum allowable control current, several performance indices were extracted from the experimental responses, as summarized in Table~\ref{tab:closed-loop-indices}. In particular, the residual RMS error $e_{\mathrm{RMS,res}}$ quantified the oscillation floor in the final part of the acquisition. In addition, the attenuation of the dominant oscillatory mode, $A_{f_n}$, and the integral of the radius signal, $J_r=\int r(t)\,dt$, were considered to jointly evaluate frequency-domain suppression and cumulative oscillation energy. The time $T_{50}$ was also reported to characterize the initial damping speed.
        \begin{table}[!t]
        \begin{center}
        \caption{Closed-loop performance indices for different maximum control currents.}
        \label{tab:closed-loop-indices}
        \begin{tabular}{|c|c|c|c|c|}
        \hline
        $I_{\max}$ [A] & $T_{50}$ [s] & $e_{\mathrm{RMS,res}}$ [mm] & $A_{f_n}$ [dB] & $J_r$ [mm s] \\
        \hline
        0.0   & 11.10 & 1.058  & --    & 19.47 \\
        \hline
        0.1   & 6.71  & 0.552  & 3.28  & 14.53 \\
        \hline
        0.2   & 4.06  & 0.0458 & 7.34  & 8.81 \\
        \hline
        0.3   & 2.42  & 0.0448 & 17.37 & 5.84 \\
        \hline
        0.4   & 1.87  & 0.0605 & 25.63 & 4.49 \\
        \hline
        0.6   & 1.01  & 0.1187 & 31.11 & 3.27 \\
        \hline
        0.8   & 0.70  & 0.1752 & 20.14 & 3.40 \\
        \hline
        1.0   & 0.31  & 0.2051 & 22.08 & 3.98 \\
        \hline
        \end{tabular}
        \end{center}
        \end{table}
        The results highlight a clear trade-off between damping speed and steady-state residual oscillation. As expected, increasing $I_{\max}$ yields a faster initial collapse of the pendulum motion, as shown by the monotonic reduction of $T_{50}$ from $6.71$~s at $0.1$~A to $0.31$~s at $1.0$~A. However, the same increase in actuation authority also raises the residual oscillation level, with $e_{\mathrm{RMS,res}}$ growing from about $0.045$~mm at $0.2$--$0.3$~A to $0.205$~mm at $1.0$~A. The attenuation and cumulative-energy indices indicate an optimal compromise at intermediate current values. In particular, $I_{\max}=0.6$~A provides the highest attenuation of the dominant oscillatory mode ($31.11$~dB) and the lowest cumulative oscillation index ($J_r=3.27$~mm\,s). Conversely, lower current values lead to cleaner steady-state behavior but significantly slower damping, whereas higher current limits ($0.8$--$1.0$~A) further accelerate the initial transient at the expense of a markedly higher residual oscillation floor. These results indicate that the maximum control current should not be selected solely to maximize damping aggressiveness. Instead, it must be tuned as a design parameter that balances transient suppression against steady-state measurement quality. For the considered prototype, the range between $0.4$~A and $0.6$~A appears to provide the most favorable trade-off.

\section{Conclusion}
    This paper presented an active electromagnetic damping strategy for a high-precision pendulum-based inclinometer and validated it on a dedicated experimental prototype. Rather than focusing on the controller alone, the main contribution lies in an integrated framework that combines electromagnetic actuator design, control-oriented modeling of the spherical pendulum, observer-based feedback, and constrained force allocation to generate, in real time, a feasible planar damping action under unilateral and saturation constraints. FEM-supported design and energy-based feasibility analysis further showed that a six-coil layout is required to satisfy the damping objective over the full pendulum workspace. Based on this framework, the closed-loop strategy was implemented and experimentally assessed for different current limits. The results demonstrated a marked improvement in dynamic behavior, with attenuation of the dominant oscillatory mode up to 31.11 dB. Moreover, the transient damping time \(T_{50}\) was reduced from 11.10 s in the uncontrolled case to 1.01 s at \(I_{\max}=0.6\) A, while more aggressive actuation further shortened the initial transient down to 0.31 s at 1.0 A. Intermediate current limits, approximately 0.4 - 0.6 A for the considered prototype, provided the most favorable compromise between rapid suppression of the disturbance transient and preservation of steady-state measurement quality. The residual oscillation floor at the end of the transient should not be interpreted solely as an intrinsic limitation of the proposed damping principle, since it may also reflect practical factors such as measurement noise, limited sensing bandwidth, indirect velocity estimation, and residual model mismatch. In addition, the control action is intentionally limited near the rest position, where velocities are very small, to avoid perturbing the equilibrium point and preserve static measurement accuracy. Despite these effects, the proposed system still achieved substantial oscillation attenuation and a drastic reduction of the transient duration on the real prototype. Overall, the results show that active electromagnetic damping can substantially improve the dynamic usability of pendulum-based inclinometers without resorting to contact-based damping methods. Beyond the specific prototype considered here, the proposed methodology provides a solid basis for the design and implementation of precision gravity-referenced sensing systems operating in mechanically disturbed environments.


\begin{IEEEbiography}[{\includegraphics[width=1in,height=1.25in,clip,keepaspectratio]{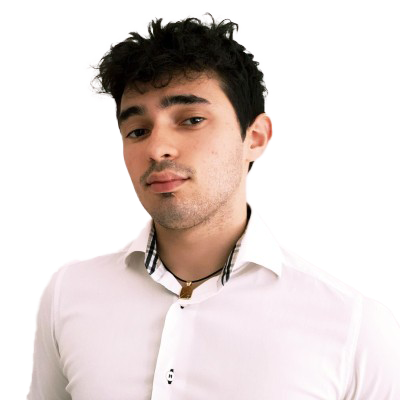}}]{Fernando Capes}
received the M.Sc. degree in Electronic Engineering from the University of Applied Sciences and Arts of Southern Switzerland (SUPSI), Lugano, Switzerland, in 2024. He is currently an Electrical Engineer with ABB, working in the field of simulation and power electronics systems. His interests include power electronics, control systems, system-level simulation, and testing methods for industrial electrical systems.
\end{IEEEbiography}

\begin{IEEEbiography}[{\includegraphics[width=1in,height=1.25in,clip,keepaspectratio]{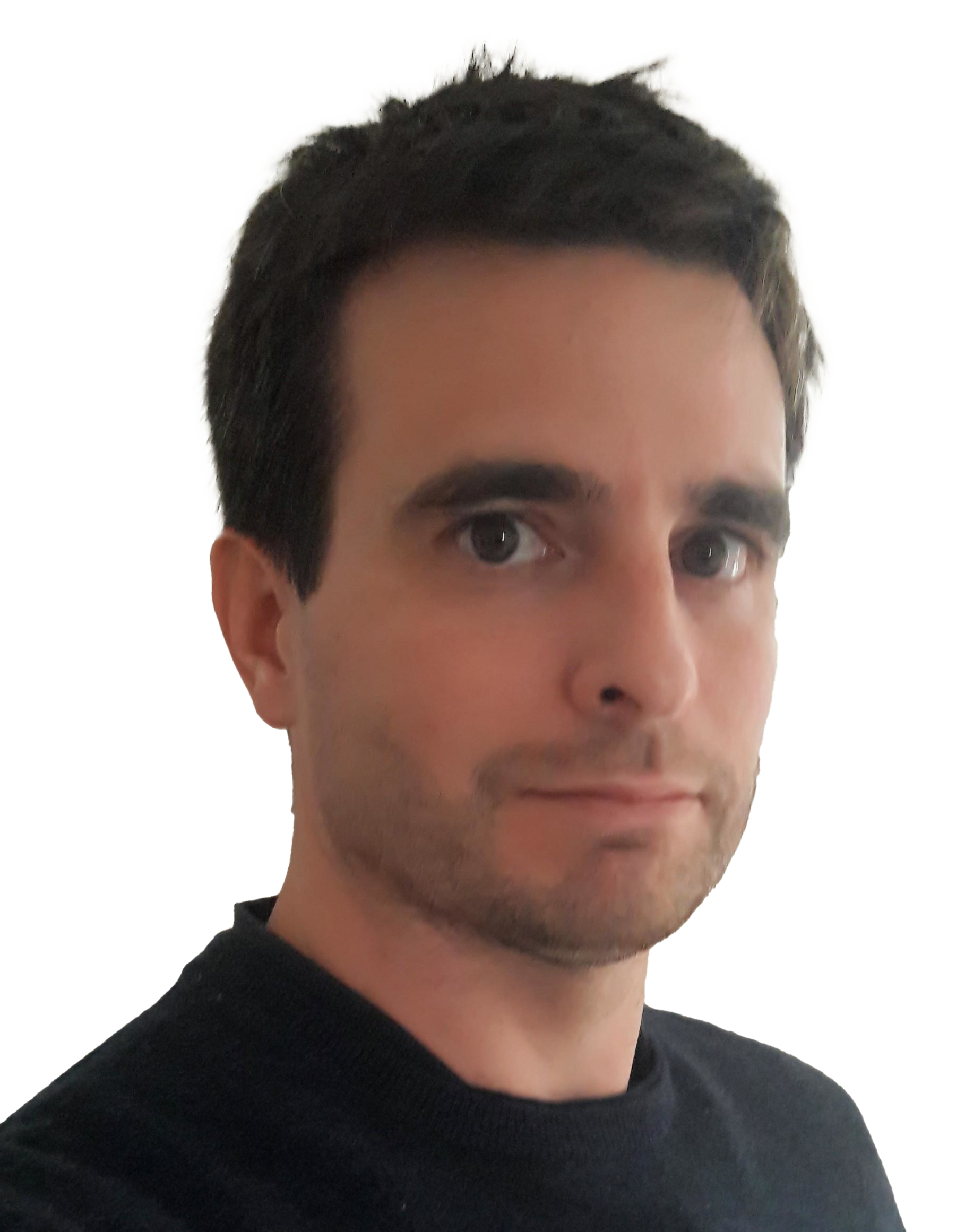}}]{Mikael Andreas Bianchi}
received the Dr. Ing. degree from ETH Zurich, Zurich, Switzerland. He is currently a Senior Researcher and Head of the Mechatronic Systems Scientific Area at the Institute of Systems and Applied Electronics, University of Applied Sciences and Arts of Southern Switzerland (SUPSI), Lugano, Switzerland. His research interests include control theory and applications, electrical machine control, precision positioning, system identification, embedded control, and modeling and validation of dynamic systems.
\end{IEEEbiography}

\begin{IEEEbiography}[{\includegraphics[width=1in,height=1.25in,clip,keepaspectratio]{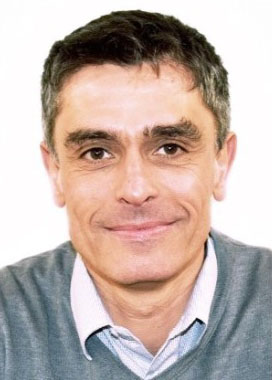}}]{Roberto Gardenghi}
is an Adjunct Professor in optoelectronic systems and applied photonics with the University of Applied Sciences and Arts of Southern Switzerland (SUPSI), Lugano, Switzerland. He is also Head of the Applied Photonics and Optoelectronics Scientific Area at the Institute of Systems and Applied Electronics. His research and professional interests include optoelectronic systems, applied photonics, sensing technologies, and measurement systems for industrial and scientific applications.
\end{IEEEbiography}

\begin{IEEEbiography}[{\includegraphics[width=1in,height=1.25in,clip,keepaspectratio]{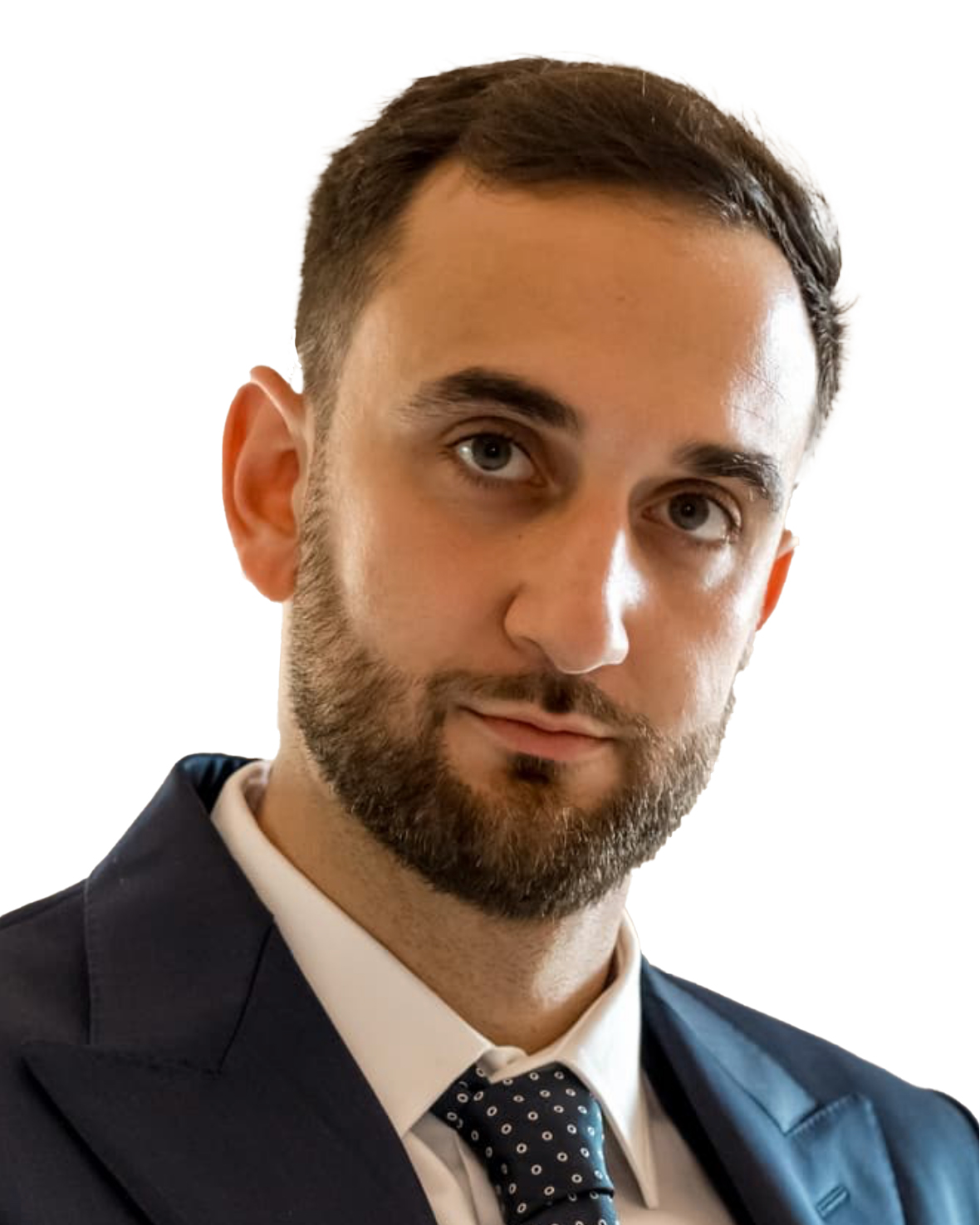}}]{Manuel Alò}
received the M.Sc. degree in electronic engineering, with a specialization in industrial technologies for precision systems, from the University of Applied Sciences and Arts of Southern Switzerland (SUPSI), Lugano, Switzerland. He is currently a Researcher with the Mechatronic Systems Laboratory, Institute of Systems and Applied Electronics, SUPSI. His research interests include high-precision mechatronic systems, control, sensing, estimation, machine vision, and advanced algorithms for industrial automation.
\end{IEEEbiography}

\vfill

\end{document}